\documentclass[preprint,showpacs,superscriptaddress]{revtex4-1}
\usepackage{graphicx}
\usepackage{amsmath}
\usepackage{amssymb}
\usepackage{color}
\usepackage{bm}
\usepackage{braket}

\begin{document}

\title{Public opinion by a poll process: model study and Bayesian view}

\author{Hyun Keun Lee}
\affiliation{Department of Physics, Sungkyunkwan University, Suwon 16419, Korea}
\author{Yong Woon Kim}
\email{y.w.kim@kaist.ac.kr}
\affiliation{Graduate School of Nanoscience and Technology, Korea Advanced Institute of Science and Technology, Deajeon 34141, Korea}
\affiliation{Department of Physics, Korea Advanced Institute of Science and Technology, Deajeon 34141, Korea}

\date{\today}
\begin{abstract}
We study the formation of public opinion in a poll process where the current score is open to public. The voters are assumed to vote probabilistically for or against their own preference considering the group opinion collected up to then in the score. The poll-score probability is found to follow the beta distribution in the large polls limit. We demonstrate that various poll results even contradictory to the population preference are possible with non-zero probability density and that such deviations are readily triggered by initial bias. It is mentioned that our poll model can be understood in the Bayesian viewpoint.
\end{abstract}

\maketitle

\section{introduction}
\label{intro}
Public opinion is a collective attitude of individuals on social or political issues, and its formation is governed by the interactions among individuals and the influence of mass media~\cite{po,po2}. As some characteristics to reach consensus are regarded to have an analogy with the magnetization of spin alignment, the spin systems have been providing the research tool in studying the public opinion formation~\cite{sv0,sv,sv2,sv3,sv4,svp,svp2,svp3}. Also, the heterogeneity of the interaction structure was found to play a significant role as the interaction networks with hubs of many connections show the distinctive opinion formation not observed in the earlier works~\cite{svn,svn2}. Recently, an empirical study is also performed to examine the opinion dynamics with the real data available in the social media~\cite{empstd}.

In the public opinion formation, the most interactions are considered local except for a few exceptions like that of hub node, for example, because of the practical restrictions of the physical entity interaction. Recently, the restriction is however reduced a lot in the progress of communication technology. The vast amount of information by anonymous agents is spread across the world instantly through various media and readily influences individuals. Furthermore, opinions on a specific issue are easily gathered, publicized, and again affects the opinion of the other individuals who might not have a firm opinion or belief. The influenced opinion then propagates again and by itself also has an impact on the opinion of others. Therefore, in the environment of a fast and wide range of communications, opinion formation can be a consequence of a chain reaction triggered by pre-formed opinion, as observed in information cascade~\cite{anderson}. This is as a matter of fact what happens in many web-based opinion propagations on social/political issues, new products, and fake news~\cite{fn}.

In this paper, we investigate public opinions in a poll model, proposed focusing upon the effect of prior opinion. In our model, the poll score is released every time a new vote comes in, providing a prior opinion to the next voter. Voters are assumed to vote stochastically for or against their own preferences with a probability that depends on the poll score at the moment of voting and the self-assurance about the preference (or faith).
The probability distribution of the poll score is found to follow the beta distribution in the large polls limit. It is demonstrated that various poll results even contradictory to the population preference are possible with non-zero probability density and that such deviations are readily triggered by initial bias. A Bayesian interpretation of our model is finally proposed.

This paper is organized as follows. In Sec.~\ref{ms}, we propose a poll model where individual stochastically votes for or against her/his preference considering the poll score, and then solve it analytically in Sec.~\ref{s}. In Sec.~\ref{pop}, a few interesting properties of the solution in the aspect of opinion formation are illustrated. In Sec.~\ref{bi}, we discuss that our model can be understood in the Bayesian viewpoint. A few remarks are finally added in Sec.~\ref{d}.

\section{model}
\label{ms}

We consider a poll where the voter has two options, $A$ or $B$. The poll score is counted per each vote, and is open to the public. Now suppose a voter who prefers $A$ is about to poll while knowing that the option $A$ scores $n$ out of the previous $t$ votes. We here assume that the voter partially complies with the group opinion $n/t$ representing the average preference of the previous voters to the option $A$. This consideration weakens the voter's preference to $A$. In order to model this feature stochastically, we introduce a probability $\sigma$ between $n/t$ and $1$, with which the voter votes for $A$. With the complementary probability $1-\sigma$, the voter votes for $B$ instead.

To be specific, we suggest an interpolation between $n/t$ and $1$ for the probability $\sigma$:
\begin{equation}
\label{sg}
\sigma(n,t,k)
=\frac{t\times (n/t) + k\times 1}{t+k}
=\frac{n+k}{t+k}~,
\end{equation}
where the weighting factor $t$ is motivated by the number of the previous voters giving $n/t$ and the other weighting factor $k$ represents a degree of the self-assurance of the voter's own preference. As $k$ represents the self-assurance, $k>0$ is assumed.

It is instructive to see a few extreme cases of Eq.~(\ref{sg}). If $k\rightarrow \infty$ or if $n=t$, which corresponds to either the case when the voter has definite preference undisturbed or the case when the group opinion unanimously coincides with the voter's preference, the probability becomes maximized as $\sigma=1$. In the absence of previous poll results ($t=0$ and hence $n=0$), voters also vote for their preferences with probability, $\sigma=1$. If $k\approx 0$, which is the case that the voter has little confidence about the preference, $\sigma$ is almost given by the group opinion $n/t$ collected up to then.

Recall here that $\sigma$ given in Eq. (\ref{sg}) is for a voter whose preference is $A$. If the preference of a voter is $B$, the voter votes for the option $B$ with a probability $\sigma(t-n,t,k)$. We remark that no individual-dependence is considered in $k$, for simplicity and tractability. Thus $k$ also represents a collective property of the population.

\section{Poll-score distribution}
\label{s}
Suppose that $f$ and $1-f$ fractions of the total population have the preference $A$ and $B$, respectively, and assume that individuals vote sequentially in random order. Then, according to the voting rule suggested above, the probability that the poll score of the option $A$ increases by a voter is given by
\begin{eqnarray}
\label{pint}
p_{A}(t,n)&\equiv& f\sigma(n,t,k)+(1-f)[1-\sigma(t-n,t,k)] \nonumber \\
&=&\frac{n+kf}{t+k}~.
\end{eqnarray}
The first and the second term are the increments made respectively by a voter having preference $A$ and a voter who has the preference $B$ but votes for $A$. Similar consideration leads to the probability that the poll score of the option $B$ increases by a voter,
\begin{eqnarray}
\label{pintb}
p_{B}(t,n)&\equiv& f[1-\sigma(n,t,k)]+(1-f)\sigma(t-n,t,k) \nonumber \\
&=&1-p_{A}(t,n)~.
\end{eqnarray}
This is same as the probability that the poll score of the option $A$ remains unchanged.

Calculation relevant to our interest can proceed more transparently with a notation $\ket{t,n}$, representing the poll state of score $n$ of option $A$ after $t$ votes, and two linear operators $\hat A$ and $\hat B$, defined as
\begin{eqnarray}
\label{A}
\hat A\ket{t,n} &=&
p_{A}(t,n)\ket{t+1,n+1}\\
\label{B}
\hat B\ket{t,n} &=&
p_{B}(t,n)\ket{t+1,n}~.
\end{eqnarray}
The $\hat A$ operation on $\ket{t,n}$ increases both the total vote number and the score for $A$ by one and gives the probability $p_{A}$ as the proportional coefficient. The $\hat B$ operation increases only the total vote number, and the resulting state $\ket{t+1,n}$ is multiplied by the coefficient, $p_{B}$, the probability of its occurrence. Adding Eqs.~\eqref{A} and \eqref{B}, we have $\hat V\equiv (\hat A + \hat B)$ acting on $\ket{t,n}$ to yield
\begin{equation}
\label{evol}
\hat V \ket{t,n} =
p_{A}(t,n)\ket{t+1,n+1}+ p_{B}(t,n)\ket{t+1,n}~,
\end{equation}
which is a useful formula in the probabilistic description of the voting result. For example, $\hat V^{2}\ket{t,n}$ is given by a linear combination of $\ket{t+2,n}, \ket{t+2,n+1}$, and $\ket{t+2,n+2}$, of which each coefficient is the probability for 
the multiplied poll state to appear. In this way, repeated application of $\hat V$ leads to
all possible poll states with their own probabilities.

Let $\ket{t_0,n_0}$ with integer $t_0 \ge n_0 \ge 0$ be the initial condition of the poll. We do not impose $t_0=0$ to emulate a possible initial guide (or bias) by a few experts' opinion or pre-poll voting on the issue, for example. All possible poll states that occur after $t$ votes can be generated by applying $\hat V$ $t$ times to the initial state
$\ket{t_{0},n_0}$ as
\begin{equation}
\label{last}
\hat V^t\ket{t_0,n_0}=\sum_{n=0}^{t}P(t,n)\ket{t_0+t,n_0+n}~.
\end{equation}
Here $P(t,n)$ is the probability that the option $A$ acquires $n$ votes from $t$ voters, which is the central quantity of our interest.

In deriving $P(t,n)$, we first check whether one can use the binomial expansion for $\hat V^t = (\hat A+\hat B)^t$ or not. For an arbitrary $\ket{t,n}$, using Eqs.~\eqref{A} and \eqref{B}, we compare
\begin{eqnarray}
\label{AB}
\nonumber
\hat A \hat B\ket{t,n} &=& p_{B}(t,n)\hat A |t+1,n\rangle \\
&=& p_{B}(t,n)p_{A}(t+1,n)|t+2,n+1\rangle
\end{eqnarray}
with
\begin{eqnarray}
\label{BA}
\nonumber
\hat B\hat A \ket{t,n}&=& p_{A}(t,n)\hat B \ket{t+1,n+1}\\
&=&p_{A}(t,n)p_{B}(t+1,n+1)\ket{t+2,n+1}~.
\end{eqnarray}
Here, one may use Eqs.~\eqref{pint} and \eqref{pintb} to find
\begin{equation}
\label{rwid}
p_B(t,n)p_A(t+1,n)=p_A(t,n)p_B(t+1,n+1)
=\frac{(n+kf)(t-n+k(1-f))}{(t+k)(t+1+k)}~.
\end{equation}
Equation~\eqref{rwid} shows that $\hat A$ and $\hat B$ commute with each other since we consider an arbitrary $\ket{t,n}$. Thus, regardless of the past, the probability of a vote for $A$ and then for $B$ is equal to the probability of a vote for $B$ and then for $A$. Hence, the binomial expansion,
\begin{equation}
\label{bi}
\hat V^t = \sum_{n=0}^{t}
\frac{t!}{(t-n)!n!}
\hat B^{t-n} \hat A^n~,
\end{equation}
can be utilized. Introducing $c(t,n)$ as $\hat B^{t-n} \hat A^n \ket{t_0,n_0}=c(t,n)\ket{t_0,n_0}$, we write
\begin{equation}
\label{ptnc}
P(t,n)=
\frac{t!}{(t-n)!n!} c(t,n)
\end{equation}
and obtain $c(t,n)$ through repeated application of Eqs.~\eqref{A} and \eqref{B}. A little algebra leads to
\begin{eqnarray}
\label{c}
c(t,n) &=& \prod_{j=1}^{t-n}
\frac{t_0+j-1-n_0+k(1-f)}{t_0+n+j-1+k}
\prod_{j=1}^{n} \frac{n_0+j-1+kf}{t_0+j-1+k} \nonumber \\
&=&
\frac{\Gamma(t_0+t-n-n_0+k(1-f))\Gamma(t_0+n+k)}
{\Gamma(t_0-n_0+k(1-f))\Gamma(t_0+t+k)}
\nonumber \\
&& \times
\frac{\Gamma(n_0+n+kf)\Gamma(t_0+k)}
{\Gamma(n_0+kf)\Gamma(t_0+n+k)}~,
\end{eqnarray}
where $\Gamma(z)$ is the gamma function~\cite{Arfken}, and its property $\Gamma(z+1)=z\Gamma(z)$ is used to obtain the second equality. Inserting Eq.~\eqref{c} into Eq.~\eqref{ptnc} and rearranging terms, we reach
\label{ptn0}
\begin{eqnarray}
P(t,n)&=&
\frac{\Gamma(\alpha+\beta)}{\Gamma(\alpha)\Gamma(\beta)}
g(t,1,t_0+k)g(n,\alpha,1)g(t-n,\beta,1)~,
\end{eqnarray}
where the parameters $\alpha$ and $\beta$ are defined as
\begin{equation}
\label{ab}
\alpha = n_0+kf~~{\rm and}~~\beta=t_0-n_0+k(1-f),
\end{equation}
and $g(z,a,b)$ is the ratio of two gamma functions: $g(z,a,b)=\Gamma(z+a)/\Gamma(z+b)$.

Because in most of the poll performed the number of votes is usually large, it is more meaningful to find the behavior of $P(t,n)$ for $t\gg 1$. Moreover, unless either of the option $A$ and the option $B$ is absolutely supported such that $n \sim {\cal O}(1)$ and $t-n\sim {\cal O}(1)$, which is actually the case of polling on a subtle and controversial issue, $n$ and $t-n$ can also be assumed large numbers. Under this considerations, using that the gamma function ratio follows an asymptotic behavior, $g(z,a,b)\approx z^{a-b}$ for relatively large $z$~\cite{Arfken}, we find that final voting results are well characterized by
\begin{eqnarray}
\label{ptna}
P(t,n) &\approx&
\frac{\Gamma(\alpha+\beta)}{\Gamma(\alpha)\Gamma(\beta)}
t^{1-t_0-k} n^{\alpha-1} (t-n)^{\beta-1}~.
\end{eqnarray}
With $r\equiv n/t$, the right hand side of Eq.~\eqref{ptna} is rewritten as $(\frac{1}{t})\frac{\Gamma(\alpha+\beta)}{\Gamma(\alpha)\Gamma(\beta)} r^{\alpha-1}(1-r)^{\beta-1}$. In the large $t$ limit, $r$ becomes continuous and the infinitesimal quantity $dr=dn/t=1/t$ appears. Therefore, one may read the probability {\it density} function 
\begin{equation}
\label{pxc}
p(r)=
\frac{\Gamma(\alpha+\beta)}{\Gamma(\alpha)\Gamma(\beta)} r^{\alpha-1} (1-r)^{\beta-1}~
\end{equation}
on $0<r<1$, which satisfies $\int_{n_1/t}^{n_2/t} dr p(r) = \sum_{n=n_1}^{n_2}P(t,n)$ for any $n_1$ and $n_2$ in the large $t$ limit. Interestingly, $p(r)$ is the well-known beta distribution~\cite{distr}, of which shape parameters $\alpha$ and $\beta$ are given by Eq.~(\ref{ab}). The various statistical properties of $r$ are therefore available from what is already known for the beta distribution.

Finishing this section, we add two remarks. The validity of $p(r)$ [Eq.~\eqref{pxc}] is restricted when one of $\alpha$ and $\beta$ vanishes. In this case, the beta distribution is ill-defined because of $\Gamma(0)$. Instead, the limiting behavior of $p(r)$ is compatible
with the model result, as follows. $\alpha(=n_0+kf)=0$ appears if $f=n_0=0$. In this situation, Eqs.~\eqref{pint} and \eqref{pintb} shows that the score of $B$ always increases to give $P(t,0)=1$ for all $t$. This observation is compatible with the associated limiting behavior, $\lim_{\alpha\rightarrow 0} p(r)=\delta(r)$, in that $\int_0^\epsilon dr \delta(r)=1$ for any fixed $\epsilon > 0$. That is, $\alpha=0$ case can be still understood with $p(r)$ in the $\alpha \rightarrow 0$ limit. Similarly, $\beta(=t_0-n_0+k(1-f))=0$ by $f=1$ and $n_0=t_0$ trivially gives $P(t,t)=1$ that is understandable with $\lim_{\beta\rightarrow 0}p(r)=\delta(r-1)$. In our model, $\alpha=\beta=0$ is not the case because $k>0$. We also remark that our model can be viewed as a generalization of the P\'{o}lya's urn~\cite{polya} well-known in statistics community.
Our model gives the P\'{o}lya's urn process when $t_0=0$ case is excluded and the self-assurance part is removed with $k=0$.

\section{public opinion property}
\label{pop}
Now we discuss the behaviors of the poll score distribution, $p(r)$, given as Eq.~\eqref{pxc}, and how the poll results reflect the preference of the population. The average of $r$ and its variance are, respectively, given by
\begin{equation}
\label{ar}
\langle r \rangle
= \frac{\alpha}{\alpha+\beta}
= \frac{t_0\rho_0+kf}{t_0+k}
\end{equation}
and
\begin{equation}
\label{v}
\langle
\delta r^2
\rangle
= \frac{\alpha\beta}{(\alpha+\beta)^2(\alpha+\beta+1)}
= \frac{\langle r \rangle (1-\langle r \rangle)}{t_0+k+1}~,
\end{equation}
where $\rho_0 \equiv n_0/t_0$ and $\delta r^2 \equiv (r-\langle r \rangle)^2$. Note that the average $\langle r \rangle$ is not necessarily $f$ and is given by the interpolation between $\rho_0$ and $f$ with weights $t_0$ and $k$, respectively. As a result, a finite initial score of $t_0$ and $n_0$ can make the average of poll results different from $f$ representing the preference of the population. Moreover, the variance remains finite even though $\langle r \rangle$ is the average of the infinite number of bounded random variables. These observations are attributed to the fact that the score increment in each vote is {\it not} an independent and identically distributed random variable but rather depends on the poll score up to then [see Eq.~\eqref{pint}]. In addition, the variance decreases for $t_0$, which implies the poll result could be controlled in an efficient way by increasing $t_0$. A more detailed discussion will follow.

\begin{figure}
\includegraphics*[width=\columnwidth]{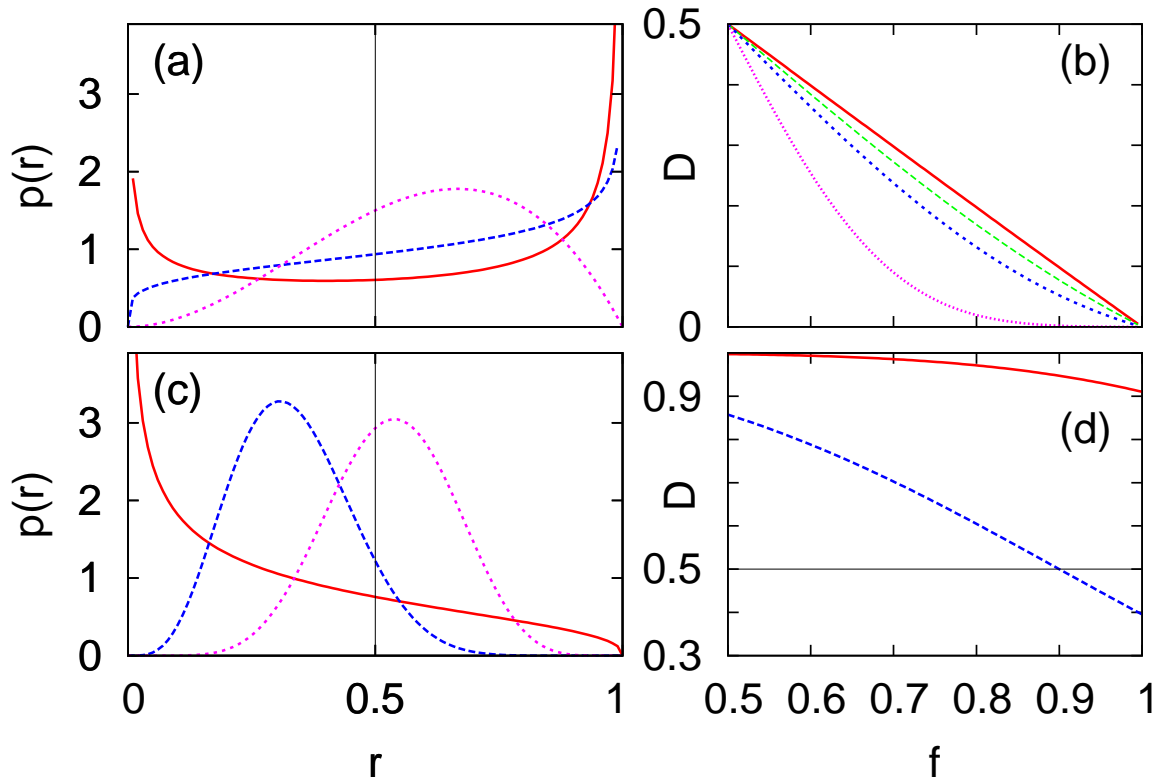}
\caption{(a) The probability density function $p(r)$ [Eq.~\eqref{pxc}] when $f=0.6$ and $t_0=0$ with $k=1$ (solid), 2 (dashed), 5 (dotted) respectively. (b) The distortion probability $d(p)$ (see text) when $f=0.6$ and $t_0=0$ with $k=0.2/1/2/10$, respectively, from top to bottom. (c) $p(r)$ with $f=0.6,k=t_0=1,n_0=0$ (solid) and $t_0=10, n_0=0/3,f=1,k=5$ (dashed/dotted). (d) $D$ as a function of $f$ with $t_0=10, n_0=0/3,k=5$ (solid/dashed).}
\label{diag}
\end{figure}

Let us first consider the case of $t_0=0$, i.e. with no initial bias. One of the interesting features of the beta distribution is that, when the shape parameters, $\alpha$ and $\beta$, are smaller than $1$, the distribution shows singular peaks at the boundaries, $r=0$ and $r=1$, while the average lies between the two peaks [the solid curve in Fig.~\ref{diag}(a)].
The two peaks at the boundaries indicate that either of the extreme poll results is likely. The condition of $\alpha,\beta<1$ basically requires $t_0=0$, and thus it is equivalent to $k < 1/\max(f,1-f)$: the boundedness of $k$ as small value suggests such an extreme poll result can be expected in a society of weak self-assurance. Note that the two peaks do imply not a polarization of the public opinion but a probabilistic bifurcation (or herding) by random voting scores at early stages.

As $k$ increases to hold $1/\max(f,1-f) < k < 1/\min(f,1-f)$, $\alpha>1$ and $\beta<1$ follows when $f>1/2$ or, otherwise, $\alpha<1$ and $\beta>1$ when $f<1/2$. Then one of the peaks disappears [the dashed curve in Fig.~\ref{diag}(a)]. Increasing $k$ further up to $k > 1/\min(f,1-f)$, the remaining peak also disappears to give a centered distribution whose maximum is in $0<r<1$ [the dotted curve in Fig.~\ref{diag}(a)]. This way, the tendency of herding to the extreme(s) is weakened as $k$, a self-assurance, increases. In any cases with $t_0=0$, $\langle r \rangle = f$ [Eq.~\eqref{ar}] implies the poll results capture the population preference, on average. However, since the variance remains finite even in the infinite population limit [Eq.~\eqref{v}], it is likely that the statistical properties of the poll results are ill-represented by the average alone. A poll result of large deviation from the population preference will not be rare even though the population is infinite.

Suppose, for instance, that an issue is determined by the majority rule with poll result. Then there is a non-zero probability of {\it distortion} in a sense that the issue will be settled with a final decision opposite to the major preference of the population. For $f>1/2$, the distortion probability is measured as $D = \int^{1/2}_{0} dr \, p (r)$, the area of $p(r)$ below $r=1/2$, while for $f<1/2$, it is given by the area above. In Fig.~\ref{diag}(b), we show the distortion probability $D$ as a function of $f$ for various $k$, which monotonically decreases from $1/2$ to $0$ as the population preference $f$ increases. For a given $f$, $D$ becomes larger for smaller $k$ and approaches $1-f$ as $p(r) \rightarrow f\delta(r)+(1-f)\delta(r-1)$ in the $k \rightarrow 0$ limit when $t_0=0$. This indicates the distorted decision is reached with higher probability in a society of weak self-assurance.

We next consider the initial bias (or intervene) case of $t_0 \ge 1$ in which the average is not kept $f$ unless $\rho_0 = f$. This may result in the various distortion depending on the choice of $t_0$ and $n_0$. For example, if $t_0\ge 1$ and $n_0=0$ are considered, the same $f$ and $k$ used in Fig.~\ref{diag}(a) give the shape parameters, $\alpha < 1$ and $\beta > 1$ [Eq.~\eqref{ab}], leading to a distribution peaked at $r=0$ [the solid curve in Fig.~\ref{diag}(c)]. An interesting point here is that the distortion probability can be greater than $1/2$: it is more probable that the poll results indicate the opposite to the preference of the population. We below call such a dominant occurrence of the distortion as the {\it reversal}. The reversal can take place in various ways depending on the combination of $t_0, n_0, f$, and $k$. The dashed curve in Fig.~\ref{diag}(c) is another example of the reversal while the dotted one therein is not. Then, a rising question is when the reversal occurs.

The condition in which the reversal occurs corresponds to $D > 1/2$. Observing Eqs.~\eqref{pxc} and \eqref{ab}, one finds that the side of major distribution with respect to $r=1/2$ changes depending upon which of $\alpha$ and $\beta$ is larger. In order for the reversal to occur, $\alpha < \beta$ is therefore required when $f>1/2$, and this results in
\begin{equation}
\label{rc}
t_0 \left(\frac{1-2\rho_0}{2f-1}\right) > k~.
\end{equation}
The criterion is also given in the same form for the case of $f<1/2$. One easily checks that each set of $t_0,n_0,p$, and $k$ of the solid/dashed curves in Fig.~\ref{diag}(c) fulfills Eq.~\eqref{rc} while that of the dotted curve does not.

Equation~\eqref{rc} shows that only a finite $t_0$ can bring about the reversal even if the infinite population size is considered. This is still the case even for a population with full consensus ($f=1$) only if $t_0 > k/(1-2\rho_0)$. When $\rho_0=0$, any finite $t_0>k$ leads to the reversal, no matter how strong the consensus is. These indicate that with $t_0$ and $n_0$ implanted, probing the population preference through a poll can be unreliable. The solid curve of Fig.~\ref{diag}(d) shows a distortion probability where the reversal occurs even for $f=1$ when $t_0=10,n_0=0$, and $k=5$. The distortion probability at $f=1$ is the distribution area of the dashed curve of Fig.~\ref{diag}(c) below $r=1/2$, and it is $0.91022(3)$. Its complementary probability, the area of the other side, decays {\it exponentially} fast as $t_0$ increases. Note this is the probability that the population of full consensus of $f=1$ will win in the poll by the majority rule. The dashed curve in Fig.~\ref{diag}(c) shows the reversal is not the case after $f=0.9$ when $n_0=3$ is instead used.

\section{Bayesian interpretation}
\label{bi}
We discuss that our model [Eq.~\eqref{sg}] can be interpreted in a Bayesian viewpoint~\cite{bf}, as follows. Let $v_i=0~{\rm or}~1$ be the voting score of $A$ by $i$-th voter. From the commutativity property $\hat A\hat B = \hat B\hat A$ [see Eqs.~\eqref{AB}, \eqref{BA}, and \eqref{rwid}], it follows that
\begin{equation}
\label{exc}
p(v_1,v_2,..,v_t)=p(v_{\pi(1)},v_{\pi(2)},..v_{\pi(t)})~,
\end{equation}
where $p(v_1,v_2,..,v_t)$ is the joint distribution for any $t$ and $\pi$ is an arbitrary permutation. Then, by the de Finetti's theorem~\cite{dft} for the representation of such exchangeable random variable  $v_i$s, one can write
\begin{equation}
\label{dft}
p(V_t)=\int d\theta \theta^n(1-\theta)^{t-n} F(\theta)~,
\end{equation}
where $V_t\equiv (v_1,v_2,..,v_t)$, $n=\sum_{i=1}^t v_i$, and $F(\theta)$ is a distribution on $0<\theta<1$.

Dividing both sides of Eq.~\eqref{dft} with $p(V_t)$, one obtains
\begin{equation}
\label{bt}
p_F(\theta|V_t) \equiv {1\over{p(V_t)}}\theta^n(1-\theta)^{t-n}F(\theta)~,
\end{equation}
which is trivially normalized for the integration with respect to $\theta$. Here, one may regard $\theta^n(1-\theta)^{t-n}$ as the likelihood for the observation $V_t$, out of $t$ independent Bernoulli trials with probability $\theta$ for $1$ in each. Then, $F(\theta)$ becomes the weight of the likelihood, which is referred to the prior belief (on the distribution of $\theta$) in Bayesian approach~\cite{bf,bs}. These observations illuminates that Eq.~\eqref{bt}, basically a rewrriting of Eq.~\eqref{dft}, is the Bayesian inference on the latent variable $\theta$. That is, $p_F(\theta|V_t)$ is the conditional probability density of $\theta$, provided the observation $V_t$, from the aspect of an individual with her/his own $F(\theta)$.

If she/he preferring $A$ votes for the first time, Eq.~\eqref{dft} gives $p(V_1=v_1=1)=\int d\theta \theta F(\theta) = 1$ in that the first voter follows one's own preference. Thus, the prior belief of the voter preferring $A$ should be characterized by $F(\theta)\rightarrow \delta(\theta-1)$. A realization of such $F(\theta)$ may read with $\kappa, \epsilon > 0$
\begin{equation}
\label{pb}
F(\theta) = \frac{\theta^{
\kappa -1}(1-\theta)^{\epsilon-1}}{B(\kappa,\epsilon)}~,
\end{equation}
where $\epsilon$ approaches $0$ later and $B(x,y)=\Gamma(x)\Gamma(y)/\Gamma(x+y)$ is for normalization. In Eq.~\eqref{pb}, we use a beta distribution to keep the conjugate pair~\cite{bs,cg} between the prior $F(\theta)$ and the posterior $p_F(\theta|V_t)$, as usual in the Bayesian studies.

When the average of $\theta$ is taken for $p_F(\theta|V_t)$, Eqs.~\eqref{dft}, \eqref{bt}, and \eqref{pb} gives
\begin{eqnarray}
\label{bj}
E(\theta)&\equiv& \int d\theta \theta p_F(\theta|V_t)
= \frac{p(V_t,v_{t+1}=1)}{p(V_t)} \nonumber
\\
&=& \lim_{\epsilon\rightarrow 0}\frac{B(n+1+\kappa,t-n+\epsilon)}{B(n+\kappa,t-n+\epsilon)} = \frac{n+\kappa}{t+\kappa}~.
\end{eqnarray}
Therein, $p(V_t,v_{t+1}=1)/p(V_t)$ points out $E(\theta)$ is the probability that the score of $A$ increases in the $(t+1)$-th vote, provided $V_t$. This is the expectation of an individual preferring $A$ since the used $F(\theta)$ is Eq.~\eqref{pb}.

Interestingly, $E(\theta)$ of Eq.~\eqref{bj} is same as $\sigma(t,n,k)$ of Eq.~\eqref{sg} with $\kappa=k$. This is the case when $F(\theta)$ of our model [Eq.~\eqref{sg}] that necessarily gives $P(V_1=v_1=1)=\int \theta F(\theta) d\theta =1$ is attributed to the form of Eq.~\eqref{pb}. This observation suggests that the probability of Eq.~\eqref{sg}, with which the voter votes for the preferred option, can be understood as the Bayesian expectation on the scoring rate of Eq.~\eqref{bj}. In this sense, the voter of our model supports one's own preference only as much as she/he expects in the Bayesian viewpoint.

\section{remarks}
\label{d}
The model parameter $k$ might not be easily quantified and, by its nature, it lies in psychology realm. As $k$ plays the important role in interpreting the results of our model, it is meaningful to estimate its magnitude. A few psychological experiments~\cite{conformity} are noteworthy here; perhaps the conformity experiment by Asch is the celebrated example. It was observed that an experiment participant could not defy the wrong answer by a group of agreed people (fewer than ten), even when asked a question without ambiguity such as to identify two lines of the same length. This experiment suggests that self-assurance could be, in fact, not so solid and $k$ in our model may not exceed $10$.

As demonstrated already, the effect of the initial intervene (or bias) with $t_0 \ge 1$ is significant in the statistics of the poll scores. It effectively controls the average [Eq.~\eqref{ar}] and also the variance [Eq.~\eqref{v}], and furthermore plays an efficient role in hiding the preference of the population [Eq.~\eqref{rc}]. Here, it is informative to note that the effect is basically a consequence of the competition between $t_0$ and $k$, and that $k$ is estimated not to exceed 10 in the previous paragraph. In this regards, it seems plausible that the release of the preliminary voting score may result in a so-called public opinion manipulation and/or fabrication. This is accordant with the claim/observation that an announcement of the election score or atmosphere in the early stage may have a considerable impact on the final result~\cite{prie,bwow}. The fake-news propagation can also be understood in our model with an inappropriate initial bias which intends to distort a fact.

Apparently, the real voting behavior is much more complicated than that we have considered in this work. Our model has a considerable room for improvement, for example, including the individual difference of $k$ or implementing other voting rules depending on issues. Nonetheless, we believe, the message of our present model study that the prior-opinion bias can drive the poll result against the population preference should be considered as a reflection of reality to some degree, for nowadays the opinion formation through many web-based surveys with polling or recommendation scores becomes more frequent and influential. We anticipate that these scores that might include a fabricated count is getting influential in forming public opinion as there appears too much (so-called) information on online for individuals to discern.

\begin{acknowledgments}
This research was supported by the NRF Grant No.~2015R1D1A1A01057842 (H.K.L.) and 2017R1A2B4007608 (Y.W.K.). This work was also supported by 2016R1D1A1A09918020.
\end{acknowledgments}



\begin{thebibliography}{99}

\bibitem{po} E. Katz and P. F. Lazarsfeld, {\it Personal Influence: the Part Played by People in the Flow of Mass Communications}, (The Free Press, New Bruncwick, 1955).

\bibitem{po2} D. J. Watts and P. S. Dodds, J. Consumer Res. {\bf 34}, 441 (2007).

\bibitem{sv0} R. J. Glauber, J. Math. Phys. {\bf 4}, 294 (1963).

\bibitem{sv} P. Clifford and A. Sudbury, Biometrika {\bf 60}, 581 (1973).

\bibitem{svp} S. Galam, J. Math. Psychology {\bf 30}, 426 (1986); J. Stat. Phys. {\bf 61}, 943 (1990); Eur. Phys. J. B {\bf 25}, 403 (2002).

\bibitem{svp2} K. Sznajd-Weron and J. Sznajd, Int. J. Mod. Phys. C {\bf 11}, 1157 (2000).

\bibitem{sv2} G. Szab\'{o} and A. Szolnoki, Phys. Rev. E {\bf 65}, 036115 (2002).

\bibitem{sv3} S. Fortunato, Int. J. Mod. Phys. C {\bf 16}, 17 (2005).

\bibitem{sv4} A. C. R. Martins, Int. J. Mod. Phys. C {\bf 19}, 617 (2008).

\bibitem{svp3} N. Crokidakis and P. M. C. de Oliveira, Phys. Rev. E {\bf 92}, 062122 (2015).

\bibitem{svn} V. Sood and S. Redner, Phys. Rev. Lett. {\bf 94}, 178701 (2005).

\bibitem{svn2} K. Suchecki, V. M. Egu\'{i}luz, and M. S. Miguel, Phys. Rev. E {\bf 72}, 036132 (2005).

\bibitem{empstd} F. Xiong and Y. Liu, Chaos {\bf 24}, 013130 (2014).

\bibitem{anderson} L. R. Anderson and C. A. Holt, Am. Econ. Rev. {\bf 87}, 847 (1997).

\bibitem{fn} H. Allcott and M. Gentzkow, J. Econ. Perspect. {\bf 31}, 211 (2017).

\bibitem{Arfken} G. B. Arfken and H. J. Weber, {\it Mathematical Methods for Physicists}, 6th Edition (Elsevier, Amsterdam, 2005).

\bibitem{distr} L. Devroye, {\it Non-Uniform Random Variate Generation} (Springer-Verlag, New York, 1986).

\bibitem{polya} F. Eggenberger and G. P\'{o}lya, J. Appl. Math. Mech. {\bf 3}, 279 (1923).

\bibitem{bf} {\it Foundations of Bayesianism}, edited by D. Corfield and J. Williamson (Springer-Science+Business Media, Dordrecht/Boston/London, 2001).

\bibitem{dft} B. D. Finetti, Fund. Math. {\bf 17}, 298 (1931).

\bibitem{bs} A. Gelman, J. B. Carlin, H. S. Stern {\it et. al.}, {\it  Bayesian Data Analysis}, 3rd Edition (Chapman and Hall/CRC, Boca Raton, 2014).

\bibitem{cg} D. Draper, {\it Bayesian modeling, inference and prediction} (https://users.soe.ucsc.edu/\~{}draper/draper-BMIP-dec2005.pdf, 2005).

\bibitem{conformity} R. B. Cialdini and N. J. Goldstein, Annu. Rev. Psychol. {\bf 55}, 591 (2004); S. E. Asch, Scientific American {\bf 193}, 31 (1955).

\bibitem{prie} C. Emery, {\it Public opinion polling in Canada}, (Library of Parliament, Canada, 1994).

\bibitem{bwow} T. Bale, Representation {\bf 39}, 15 (2002).

\end{thebibliography}
\end{document}